\pdfoutput=1
\documentclass[conference,10pt]{IEEEtran}
\usepackage[utf8]{inputenc}
\usepackage[T1]{fontenc}
\usepackage{hyperref}
\usepackage{url}
\usepackage{xurl}
\usepackage{booktabs}
\usepackage{array}
\usepackage{amsmath,amsfonts}
\usepackage{newtxtext,newtxmath}
\usepackage{microtype}
\usepackage{graphicx}
\usepackage{cite}
\usepackage{placeins}
\usepackage{xcolor}
\usepackage{xspace}
\usepackage{pifont}
\usepackage{algorithm}
\usepackage{algpseudocode}
\usepackage{soul}
\usepackage{tikz}
\usepackage{eso-pic}
\usepackage[most]{tcolorbox}
\graphicspath{{figs/}}
\definecolor{figblue}{HTML}{2A78D6}\definecolor{figorange}{HTML}{EB6834}

\newtcolorbox{rqsummary}[1]{enhanced, colback=gray!5!white,colframe=gray!5!black,reset,
boxsep=2pt,left=1pt,right=1pt,top=3pt,bottom=2pt,arc=0pt,boxrule=0pt,bottomrule=0.5pt,toprule=0.5pt}
\usetikzlibrary{positioning,arrows.meta,calc,shapes.geometric,fit}
\sethlcolor{yellow}

\definecolor{rowtint}{gray}{0.94}
\definecolor{headtint}{gray}{0.88}

\newcommand{\BfPara}[1]{{\noindent\ul{\bf#1.}}\xspace}
\newcommand*\cibbox[1]{\tikz[baseline=(char.base)]{\node[shape=rectangle,fill=gray!30!white,text=black,inner sep=1.5pt] (char) {#1};}}

\newcommand{\UlPara}[1]{\noindent\cibbox{{\bf{{#1.}}}}\xspace}

\setul{0.8pt}{}
\hypersetup{
	colorlinks=true,
	urlcolor=blue!70!black,
	linkcolor=red!70!black,
	citecolor=purple,
}

\newcommand{\aicite}{\texorpdfstring{~\protect\cite{openaicodex,anthropicclaude}}{}}

\newcounter{obs}
\newcommand{\observation}[2]{\refstepcounter{obs}\label{#1}\begin{rqsummary}{}\textbf{Observation~\theobs.} #2\end{rqsummary}}

\title{How the Dirty Set Moves: Characterizing Schedule\\Reuse in Incremental Sparse Matrix Multiplication}
\author{%
\IEEEauthorblockN{Hunter Horne\IEEEauthorrefmark{1}, Eric Spencer\IEEEauthorrefmark{2}\IEEEauthorrefmark{1}, Alex Henningfield\IEEEauthorrefmark{1}, Arslan Bisharat\IEEEauthorrefmark{2}}
\IEEEauthorblockA{\IEEEauthorrefmark{1}HorneSci}
\IEEEauthorblockA{\IEEEauthorrefmark{2}Loyola University Chicago, Chicago, IL, United States of America}
\IEEEauthorblockA{\{hunter.horne, alex.henningfield\}@hornesci.com, \{espencer2, marslan\}@luc.edu}}

\begin{document}
\raggedbottom
\AddToShipoutPictureBG*{\AtPageUpperLeft{\hspace*{\dimexpr\paperwidth-0.55in-1.1in\relax}\raisebox{-0.1in}{\parbox[t]{1.1in}{\vspace{0pt}\centering
\includegraphics[height=0.3in]{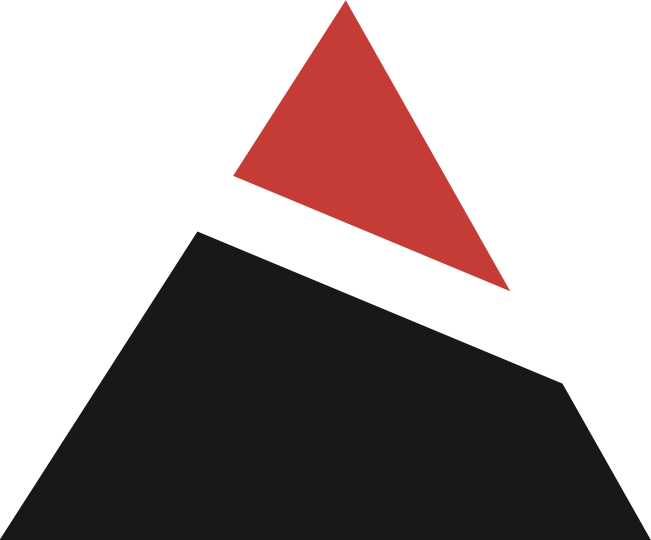}\\[1pt]
{\footnotesize\bfseries HorneSci}\\[-1pt]
{\scriptsize\url{https://hornesci.com}}}}}}
\maketitle

\begin{abstract}
Sparse matrices, in which most entries are zero, are central to scientific simulation, graph analytics and machine learning, and their multiplication with dense inputs is a core parallel kernel. In many applications the matrix stays fixed while only a few input rows change per step, so the result can be updated from those rows alone, which we call the dirty set. Sparse kernels often add a second optimization based on the inspector-executor principle, in which a tiled schedule is prepared once and reused across calls. However, when this reuse remains profitable as the dirty set moves is not well characterized. In this paper, we show that retained tiled execution rarely beats direct CSC in incremental sparse matrix multiplication, although incremental execution itself remains highly profitable. Across 21 SuiteSparse matrices and controlled patterns of motion, a direct column-wise incremental update is 2.1--8.2$\times$ faster than the best of eight full-recomputation codes, among them Intel oneMKL and PyTorch. Yet the same direct update outperforms a tiled executor that retains tile preparation and caches assembled schedules in 779 of 837 workloads. Separating reuse into per-tile preparation and cached schedules shows that the benefit comes from preparation; extra cache capacity seldom helps because a tile set rarely returns after a different one has intervened. Tiling is faster only when tiles merge many entries into each output write, a structural property that we predicted before measurement and confirmed on four held-out matrices of the same family. These results show that the structure of the matrix and the motion of the dirty set, rather than the sophistication of the schedule, determine whether reuse is profitable. We anticipate that our methodology and guidelines will help developers decide when to build schedule reuse into incremental sparse runtimes.
\end{abstract}

\section{Introduction\aicite}
\label{sec:intro}

A \emph{sparse matrix} is a matrix in which most entries are zero. Matrices from circuit simulation, power-grid analysis, structural mechanics and graph analytics routinely have millions of rows but only a few nonzeros per row~\cite{davis2011}. Sparse formats store only the nonzero entries and their positions. The compressed sparse row (CSR) format groups the nonzeros of each row, and the compressed sparse column (CSC) format groups the nonzeros of each column. Kernels that operate on these formats skip the zeros entirely, so their cost scales with the number of nonzeros rather than with the full matrix size. The product of a sparse matrix with a dense vector (SpMV) or with a dense matrix of several columns (SpMM) is therefore a core kernel in scientific computing, graph analytics and machine learning~\cite{yang2022graphblast,davis2019}.

In many applications the sparse matrix, or \emph{operator}, stays fixed while its input evolves gradually. A recommender system rescores items as users submit new ratings~\cite{harper2015}. Streaming graph systems propagate updates from a small set of modified vertices~\cite{mariappan2021}. Incremental view maintenance and differential dataflow apply the same principle to queries and iterative computations~\cite{koch2014,mcsherry2013}. In each case, a step changes only a small subset of the input rows.

For a fixed operator $A$, this structure allows an exact incremental update. Let $X_t = X_{t-1} + \Delta X_t$ hold $B$ input columns at step $t$, and let $D_t$ be the indices of the nonzero rows of $\Delta X_t$, that is, the row support of the update. Following the systems convention of marking modified data as \emph{dirty}, we call $D_t$ the \emph{dirty set}. Let $A_{D_t}$ denote the columns of $A$ indexed by $D_t$, and let $Z_t$ denote the rows of $\Delta X_t$ indexed by $D_t$. By linearity,
\begin{equation}
  Y_t = A X_t = Y_{t-1} + A_{D_t}\,Z_t .
  \label{eq:delta}
\end{equation}
The update touches only the columns of $A$ indexed by $D_t$, so its cost scales with the dirty set rather than with the input. We refer to this computation as \emph{delta-input SpMM} and to its use in place of full recomputation as \emph{delta execution}. For $B=1$ it reduces to sparse matrix by sparse vector multiplication (SpMSpV). The way $D_t$ changes from step to step is the \emph{support motion}, and the fraction of $D_t$ replaced per step is the \emph{turnover rate}.

\BfPara{Existing approaches} Sparse kernels for sparse inputs select active columns and sum their contributions~\cite{azad2017,yang2015,chen2022}. Recent designs go further and prepare the operator for repeated use. A common preparation is \emph{tiling}, which groups a small number of neighboring columns into a block called a \emph{tile}; the kernel prepares each tile once and then processes it as a unit. TileSpMSpV converts the matrix into tiles and reuses the conversion across calls~\cite{ji2022tile}. DB-SpMSpV shares blocked payloads across row and column views for changing vectors~\cite{cong2026db}. Adaptive SpMV/SpMSpV and HAM-SpMSpV select kernels from input features~\cite{li2021adaptive,xu2024ham}. Table~\ref{tab:closest} summarizes the closest designs. These designs inherit the inspector-executor principle, in which an upfront analysis is amortized over many executions that share an access pattern~\cite{vuduc2005oski,strout2018}.

\begin{table*}[t]
\centering\footnotesize
\caption{Closest sparse-input execution strategies. Rows summarize what each cited study executes, how long its prepared state lives, which inputs it adapts to at run time and which costs it reports. Entries describe the cited studies and are not performance comparisons across hardware. This study differs by varying the two lifetimes of prepared state separately and by charging setup, state and output delivery.}
\label{tab:closest}
\setlength{\tabcolsep}{3pt} \renewcommand{\arraystretch}{1.12}
\begin{tabular}{@{}>{\raggedright\arraybackslash}p{.94in}>{\raggedright\arraybackslash}p{1.11in}>{\raggedright\arraybackslash}p{1.40in}>{\raggedright\arraybackslash}p{1.31in}>{\raggedright\arraybackslash}p{1.89in}@{}}
\toprule
Method & Execution unit & Preparation lifetime & Runtime inputs & Reported cost scope \\
\midrule
Work-efficient SpMSpV~\cite{azad2017} & Selected CSC columns, output buckets & Existing CSC matrix, per-call buffers & Sparse vector indices and values & Multithreaded SpMSpV, kernel time along BFS frontiers \\
Adaptive SpMV/SpMSpV~\cite{li2021adaptive} & Eight GPU kernel paths & CSR/CSC and matrix features retained, trained selector & Current vector features, vector conversion when needed & Kernels, prediction overhead, BFS and incremental PageRank \\
TileSpMSpV~\cite{ji2022tile} & Matrix and vector tiles, push and pull & Converted matrix reused across calls & Current sparse-vector tiles and frontier & SpMSpV and BFS, conversion measured separately \\
DB-SpMSpV~\cite{cong2026db} & Dual-view blocks sharing one payload & Offline matrix conversion reused across vectors & Global block sparsity and local vector masks & GPU kernels, BFS and decoding, online metadata and storage \\
This study & Direct columns or tile sets, $B$ inputs sharing one dirty set & Operator preparation and session reuse, varied separately & Identical dirty-set streams, recorded reuse history & Matched interventions, setup-inclusive epochs, state, memory and output delivery \\
\bottomrule
\end{tabular}
\end{table*}

\BfPara{Limitations} In delta-input SpMM, the access pattern is the dirty set, and the dirty set moves. Rows leave and join $D_t$ at every step, so a schedule prepared for one step may not apply to the next. This implicit amortization assumption leads to three limitations in current practice. \ding{182}~No study has separated the parts of a prepared schedule that survive support motion from the parts that do not. \ding{183}~Current studies report kernel time and measure conversion or setup separately, so the complete cost of reuse (setup, state maintenance and output delivery) remains unknown. \ding{184}~Practitioners lack guidance on when schedule reuse is worth implementing at all.  In our measurements, a tiled executor is up to 5.75$\times$ faster than rebuilding a CSR view after every change, yet direct CSC outperforms it in 93\% of workloads with a moving dirty set. A study that compares only against the rebuild baseline would reach the opposite conclusion.

\begin{figure*}[t]
  \centering
  \includegraphics[width=\linewidth]{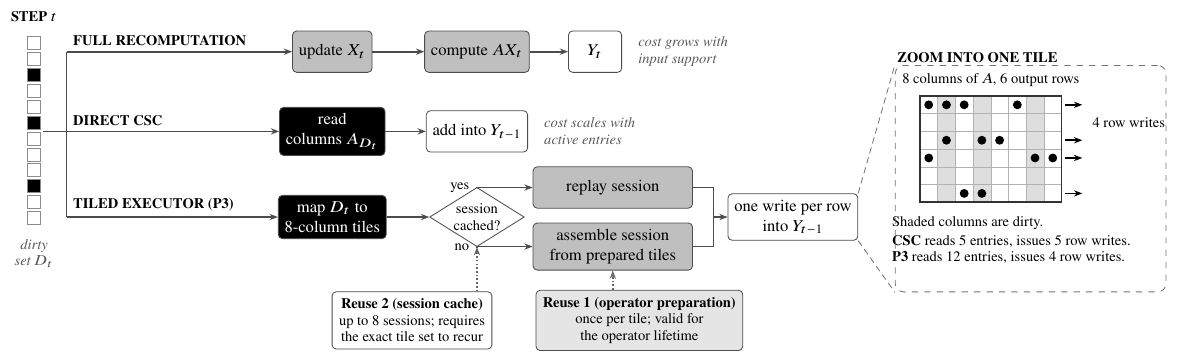}
  \caption{Execution strategies for one step of delta-input SpMM. A step changes a few input rows, the dirty set $D_t$ (black squares, left). Full recomputation (top lane) updates the input $X_t$ and recomputes the complete product $AX_t$, so its cost grows with every nonzero input row. Direct CSC (middle lane) reads only the columns of $A$ indexed by $D_t$ and adds their contribution to $Y_{t-1}$, so its cost follows the active entries. The tiled executor P3 (bottom lane) maps $D_t$ to eight-column tiles, replays a cached session when the same tile set was seen recently and otherwise assembles a new session from prepared tiles. Operator preparation (Reuse~1) is done once per tile and stays valid for the lifetime of $A$, while a session (Reuse~2) helps only when its exact tile set recurs. The inset magnifies one tile with three dirty columns (shaded). CSC reads the 5 entries in those columns and issues 5 row writes. P3 reads all 12 entries of the tile, clean columns included, but combines entries that share a row and issues only 4 row writes.}
  \label{fig:task}
\end{figure*}

\BfPara{This work} This paper presents the first systematic characterization of schedule reuse in delta-input SpMM. Our key insight is that schedule reuse has two distinct lifetimes (Fig.~\ref{fig:task}). \emph{Operator preparation} organizes the entries of each column tile once and remains valid for the lifetime of the operator, regardless of how $D_t$ moves. A \emph{session} is an assembled schedule for one exact set of tiles; it is beneficial only when that tile set recurs. We isolate each lifetime through matched interventions, that is, paired runs that change one factor while the kernel, input streams and output contract stay fixed. We then account for every cost that a deployed system would incur. The primary study times a single pinned core to isolate the per-step cost of reuse from scheduling and load imbalance; Observation~\ref{obs:e2e} extends the comparison to multicore and distributed execution.

\BfPara{Challenges} Answering this question raises three challenges. \ding{182}~Reuse must be separated from kernel, layout and input effects, so every comparison needs identical dirty-set streams and exactly one changed factor. \ding{183}~The cost of reuse is spread across setup, per-step bookkeeping, retained state and output delivery, so each term must be charged under an explicit output contract. \ding{184}~A structural explanation must be tested rather than fitted after the fact, which we address with predictions registered before timing and with matrices that the original study did not use.

\BfPara{Contributions} This paper makes the following contributions.
\begin{itemize}
\item[\ding{182}] A decomposition of schedule reuse in delta-input SpMM into operator preparation and session reuse, together with a structural work model that predicts when tiled execution can outperform direct CSC (Section~\ref{sec:background}).
\item[\ding{183}] A characterization methodology that combines generators of controlled support motion, a complete-output contract against eight full-recomputation baselines and a correctness check for every run (Section~\ref{sec:method}).
\item[\ding{184}] A characterization of 21 SuiteSparse matrices over three repeated passes and four CPU platforms, extended to a recorded application and to multicore and distributed runs (Section~\ref{sec:results}).
\item[\ding{185}] A preregistered tile-width intervention that confirms the structural source of the tiled advantage on held-out matrices, and design guidelines for incremental sparse runtimes (Observation~\ref{obs:structure} and Section~\ref{sec:guidelines}).
\end{itemize}

\BfPara{Key results} Delta execution outperforms the best of eight full-recomputation implementations by 2.1--8.2$\times$, yet direct CSC outperforms the tiled executor in 779 of 837 workloads with a moving dirty set. Tiled execution is faster only on one matrix family, whose structure compresses row writes.

\section{Background and Work Model\aicite}
\label{sec:background}

We compare four execution strategies. All receive identical dirty sets and values and produce identical results.
\begin{itemize}

\item[\ding{182}] \textbf{Full recomputation.} The input $X_t$ is updated and $AX_t$ is recomputed. We test eight implementations, namely PyTorch CSR SpMM~\cite{pytorch}, SciPy CSR and CSC products~\cite{scipy}, Intel oneMKL sparse CSR SpMM~\cite{intel2026inspector}, and four native CSC and CSR variants. The strongest native variant, which we call \emph{maintained-support} recomputation, tracks the set of nonzero input rows and visits only those columns.

\item[\ding{183}] \textbf{Direct CSC (CSC).} For each index in $D_t$, the kernel reads the corresponding column of $A$ in CSC format and adds its contribution to $Y_{t-1}$. It performs no support-dependent preparation.

\item[\ding{184}] \textbf{Exact CSR rebuild (Exact CSR).} The kernel constructs a row-major view of $A_{D_t}$ whenever the dirty set changes and reuses it while the dirty set remains unchanged.

\item[\ding{185}] \textbf{Tiled executor (P3).} Columns are grouped into tiles of width $w$ (eight by default). On first use, the executor prepares each tile by grouping its stored entries by output row, and it retains this preparation for the lifetime of the operator. It also keeps a least-recently-used cache of up to eight assembled sessions, each keyed by an exact tile set. Execution scatters the dirty values into a dense buffer, sums each row group, and issues one write per output row per tile. Algorithm~\ref{alg:step} gives one step of CSC and of P3. Two intermediate variants appear for reference. \emph{Tile CSR} (tile-cover CSR in the figures) keeps up to eight CSR views, each built for a set of tiles, and reuses any view whose tiles cover the current dirty set. \emph{Open tile} keeps the persistent tile preparation but caches only the current session.
\end{itemize}

\begin{algorithm}[t]
\caption{One step of direct CSC (lines~1--7) and of the tiled executor P3 (lines~8--26) for dirty set $D_t$ and compact update $Z_t$. $Y_i$ is row $i$ of the output, $z_j$ is the row of $Z_t$ for column $j$, $w$ is the tile width, $\mathcal{P}$ maps each tile to its prepared row groups and $\mathcal{S}$ is the least-recently-used session cache. Line~11 is session reuse and lines~13--15 are operator preparation. CSC issues one output write per stored entry (line~4). P3 first scatters $Z_t$ into a dense buffer $b$, in which clean columns of a tile stay zero, and then issues one write per row group of each tile (line~22).}
\label{alg:step}
\footnotesize
\begin{algorithmic}[1]
\Procedure{DirectCSC}{$D_t, Z_t, Y$} \For{$j \in D_t$} \For{each stored entry $(i, a)$ of column $j$ of $A$} \State $Y_i \gets Y_i + a\, z_j$ \Comment{one write per entry} \EndFor \EndFor \EndProcedure \Procedure{TiledP3}{$D_t, Z_t, Y$} \State $T \gets \{\lfloor j/w \rfloor \mid j \in D_t\}$ \Comment{tile set of $D_t$} \If{$T \in \mathcal{S}$} \State $\sigma \gets \mathcal{S}[T]$ \Else \For{$\tau \in T$ with $\tau \notin \mathcal{P}$} \State $\mathcal{P}[\tau] \gets$ entries of tile $\tau$ grouped by row \EndFor \State $\sigma \gets$ session for $T$ built from $\mathcal{P}$ \State insert $\sigma$ into $\mathcal{S}$, evict the oldest if full \EndIf \State scatter the rows of $Z_t$ into buffer $b$ of $\sigma$ \For{$\tau \in T$} \For{each row group $(i, G)$ of $\mathcal{P}[\tau]$} \State $Y_i \gets Y_i + \sum_{(j,a) \in G} a\, b_j$ \Comment{one write per row} \EndFor \EndFor \State clear the scattered rows of $b$ \EndProcedure
\end{algorithmic}
\end{algorithm}

\BfPara{Two lifetimes of reuse} The tiled executor reuses work at two granularities. Operator preparation is a per-tile cost that, once incurred, remains valid under any motion of the dirty set. Session reuse avoids reassembling the schedule for a tile set, but only when that exact tile set recurs within the cache capacity. TileSpMSpV and DB-SpMSpV convert the matrix once and reuse the conversion across calls~\cite{ji2022tile,cong2026db}, which corresponds to operator preparation in P3, and inspector-executor schedules for a fixed access pattern correspond to sessions~\cite{vuduc2005oski,strout2018}. P3 combines both forms, so our observations concern this shared reuse structure, though not the GPU implementations of those designs. Our methodology varies them independently (Observations~\ref{obs:structure} and~\ref{obs:lifetimes}).

\BfPara{Structural work model} Three quantities determine the relative cost of direct and tiled execution. The number of \emph{active entries} counts the stored entries of $A_{D_t}$. The number of \emph{covered entries} counts the stored entries in all tiles that intersect $D_t$; it is never smaller than the active count, because a tile covers clean columns as well as dirty ones. The number of \emph{row writes} counts distinct (row, tile) pairs, that is, the output updates issued after entries that share a row within a tile have been combined. Direct CSC performs work proportional to the active entries. Tiled execution performs work proportional to the covered entries plus the row writes, plus the cost of preparing previously unseen tiles. Tiled execution can therefore be faster only when \emph{write compression}, the ratio of active entries to row writes, outweighs the additional covered entries.

\BfPara{Output contracts} The comparisons among delta strategies use the \emph{delta-product contract}, under which each strategy computes $A_{D_t}Z_t$ into a zeroed output buffer. The comparison with full recomputation uses the \emph{complete-output contract}, under which every strategy returns the full $Y_t$ at every step and is charged for the state it maintains.

\BfPara{Cost accounting} We call a sequence of $L$ steps on one operator an \emph{epoch} and report two views of its cost. The \emph{steady-state} cost sums the per-step costs of cache lookups, preparation of unseen tiles and application. The \emph{setup-inclusive} cost adds one-time operator construction and per-step output resets. A strategy with a larger setup is profitable only when its total per-step savings exceed the additional setup.

\section{Characterization Methodology\aicite}
\label{sec:method}

\begin{figure*}[t]
  \centering
  \includegraphics[width=\linewidth]{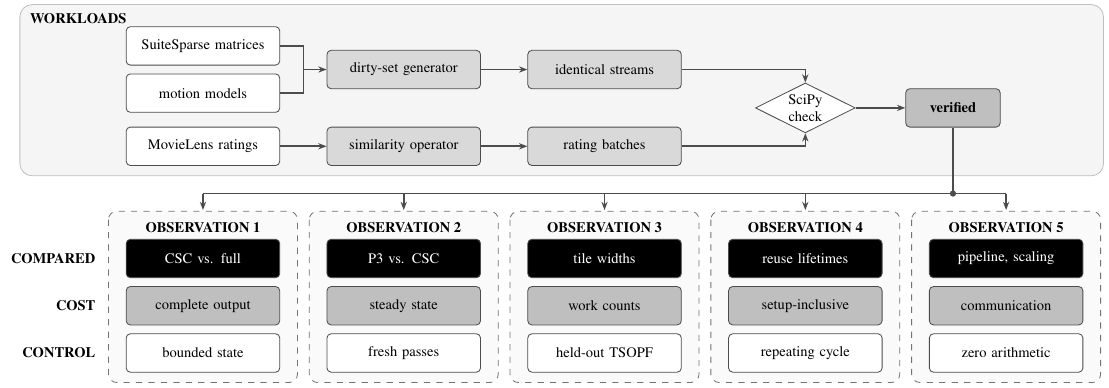}
  \caption{Characterization workflow. The workloads (top) come from two sources. Synthetic streams combine SuiteSparse matrices with two support-motion models (local drift and uniform jump) in a seeded dirty-set generator, so every strategy receives identical dirty sets and values. Recorded MovieLens 100K ratings build a fixed item-similarity operator from early ratings and replay later ratings in batches. Every session is checked against a SciPy reference, and only verified sessions reach the observations; no session failed. Each observation (bottom) lists \ding{182}~what is compared, \ding{183}~the cost charged and \ding{184}~its main control, and the matching part of Section~\ref{sec:results} gives the full design.}
  \label{fig:workflow}
\end{figure*}

Fig.~\ref{fig:workflow} summarizes the methodology. All observations draw on a single shared test set.

\BfPara{Matrices} The corpus contains 21 square matrices from the SuiteSparse collection~\cite{davis2011} in native ordering (as distributed, without reordering), with 1,723--170,998 rows and 6,511--2,600,295 nonzeros (Table~\ref{tab:corpus}). Matrices with pattern-only or near-constant values receive standard-normal values.

\begin{figure}[t]
  \centering
  \includegraphics[width=\linewidth]{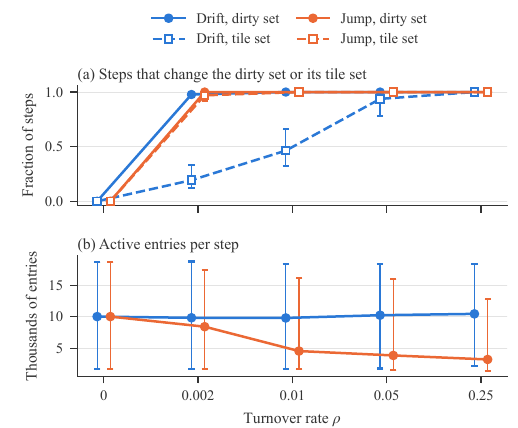}
  \caption{Realized support motion in the primary study. Markers show medians and bars show interquartile ranges over 21 matrices and five seeds at each turnover rate $\rho$, and drift and jump markers are offset horizontally for legibility. (a) Fraction of the 399 transitions of a 400-step session that change the exact dirty set (solid lines, filled markers) or its set of eight-column tiles (dashed lines, open markers). (b) Active entries, that is, stored entries in the dirty columns, per step. Under drift the dirty set changes at almost every step while its tile set changes far less often, which gives tile reuse its best opportunity. Under jump both change at nearly every step, and the active entries per step fall as $\rho$ grows.}
  \label{fig:workload}
\end{figure}

\BfPara{Support motion} Each stream starts from a dirty set grown by breadth-first search from a random column to a target of 256 rows (as few as 138 rows where the search is exhausted) and runs for 400 steps. Update values are standard-normal and identical for all strategies. At each step, a fraction $\rho$ of the dirty set is retired and replaced, with $\rho \in \{0, 0.002, 0.01, 0.05, 0.25\}$. Two generators select the replacements. \emph{Local drift} draws neighbors of surviving rows through the matrix adjacency and models activity that spreads through a network. \emph{Uniform jump} draws rows uniformly at random and models independent activity. The two models produce markedly different reuse opportunities (Fig.~\ref{fig:workload}). At $\rho=0.002$, the 104 drift sessions whose dirty set moves change the exact dirty set in a median of 390.5 of 399 transitions but change the eight-column tile set in only 78. Fig.~\ref{fig:workload} includes all 105 sessions at this rate, whose medians are 390 and 77 transitions; jump changes both in nearly every transition.

\BfPara{Platforms} Primary measurements use AWS c7i.2xlarge instances (Intel Xeon Platinum 8488C) with one pinned core and one library thread, so that all strategies compete on equal terms. Ports repeat the primary study on c7i.4xlarge (Xeon Platinum 8488C) and c7a.4xlarge (AMD EPYC 9R14), the multicore study runs on c7i.4xlarge, the distributed study on pairs of c7i.2xlarge hosts, and the setup-inclusive bounded-state control on an Intel Core i7-1165G7 mini PC. Native kernels are compiled with GCC 13.3 at \textnormal{\texttt{-O3 -march=native}}; the software stack uses Python 3.12, SciPy 1.18 and PyTorch 2.13. Unless stated otherwise, the batch width is $B=8$, the tile width is $w=8$ and the session capacity is eight.

\BfPara{Protocol} One pass covers 21 matrices, two motion models, five turnover rates and five seeds, for 1,050 sessions of 400 steps. We execute three complete passes in fresh processes~\cite{kalibera2013} on dedicated hosts; each host must show no competing load for five minutes before timing and is monitored throughout. Every strategy's output is compared with SciPy at 17 checkpoints per session, with a relative tolerance of $10^{-12}$; the maximum observed error is $2.17\times10^{-14}$. Steady-state timings exclude operator construction, which the primary passes perform once and share across strategies; the setup-inclusive control (Table~\ref{tab:setup}) instead builds the CSC operator and the P3 operator independently from identical CSR inputs and charges each its own construction. At $\rho=0$ the drift and jump labels yield identical streams, so timing both in separate sessions measures session-to-session variation, which is 2.3\% for the median matrix and at most 14.8\%. Aggregates report the median across passes of the median across seeds of the geometric mean across matrices. Of the 840 sessions with $\rho>0$, three never change their dirty set, because rounding yields zero replacements or a retired row is selected again; the remaining 837 form the \emph{moving} sessions. A result is \emph{stable} when the same strategy is faster in all three passes.

\BfPara{Methodological choices} SuiteSparse matrices are the standard corpus for sparse kernel studies~\cite{davis2011,ji2022tile}, and breadth-first growth of the initial dirty set mirrors the breadth-first search (BFS) frontiers on which SpMSpV kernels are usually measured~\cite{azad2017,ji2022tile}. Repeated runs in fresh processes follow Kalibera and Jones~\cite{kalibera2013}, and geometric means summarize ratios~\cite{fleming1986}. Two choices are new and require justification. Prior studies draw sparse inputs from graph frontiers whose motion is not controlled, which confounds turnover with structure; our generators fix the initial dirty set and vary only the turnover rate and the replacement pattern. Single-core primary timings isolate per-step cost from scheduling and load imbalance, and we report multicore and distributed runs separately.

\begin{table}[tbp]
\centering
\caption{Primary corpus of 21 square SuiteSparse matrices, used in native ordering. $N$ is the number of rows and columns and nnz the number of stored nonzeros after explicit zeros are removed. The corpus comes from power-grid, circuit, structural, fluid and other problems and spans about two orders of magnitude in size. The three TSOPF matrices at the top are the only ones on which P3 wins.}
\label{tab:corpus}
\scriptsize\setlength{\tabcolsep}{2.5pt}
\begin{tabular}{@{}lrr@{\hspace{6pt}}lrr@{}}
\toprule
Matrix & $N$ & nnz & Matrix & $N$ & nnz \\
\midrule
TSOPF\_RS\_b162\_c3 & 15,374 & 610,299 & epb2 & 25,228 & 175,027 \\
TSOPF\_RS\_b300\_c1 & 14,538 & 1,474,325 & memplus & 17,758 & 99,147 \\
TSOPF\_RS\_b39\_c30 & 60,098 & 1,079,986 & msc10848 & 10,848 & 1,229,776 \\
add32 & 4,960 & 19,848 & nasa2910 & 2,910 & 174,296 \\
bcspwr09 & 1,723 & 6,511 & olafu & 16,146 & 1,015,156 \\
bcspwr10 & 5,300 & 21,842 & raefsky3 & 21,200 & 1,488,768 \\
bcsstk18 & 11,948 & 149,090 & rajat03 & 7,602 & 32,653 \\
bcsstk25 & 15,439 & 252,241 & s3rmt3m3 & 5,357 & 207,123 \\
case9 & 14,454 & 147,972 & scircuit & 170,998 & 958,936 \\
circuit\_3 & 12,127 & 48,137 & wang3 & 26,064 & 177,168 \\
ct20stif & 52,329 & 2,600,295 & & & \\
\bottomrule
\end{tabular}
\end{table}

\section{Characterization Results\aicite}
\label{sec:results}

\UlPara{Delta versus full} Under the complete-output contract, every strategy returns $Y_t$ at every step. Table~\ref{tab:full} compares direct CSC with the full-recomputation baselines over all 1,050 sessions in three passes; all 3,150 sessions pass every check.

\observation{obs:full}{Delta execution outperforms the best of eight full-recomputation implementations by 2.1--8.2$\times$, and the margin grows with turnover.}

The best full-recomputation implementation, selected per session after timing, remains 2.10--8.19$\times$ slower than direct CSC. With frozen support, full recomputation is faster in only 12--16 of 105 sessions per pass. The margin grows because retired rows keep their nonzero values, so the nonzero input support expands over time. Full recomputation scales with this expanding support, whereas delta execution scales with the dirty set. For \textnormal{\texttt{TSOPF\_RS\_\allowbreak b300\_c1}} under jump at $\rho=0.25$, the input support reaches 12,138 rows while each step changes 256. A bounded-state control that holds the input support at 256 rows yields ratios of 2.72, 2.55 and 2.50$\times$ after 16, 64 and 400 steps, against 2.92, 3.52 and 8.46$\times$ with expanding support.

\begin{table}[t]
\centering
\caption{Full-recomputation cost relative to direct CSC under the complete-output contract, in which every strategy returns $Y_t$ at every step. Entries are ratios of 400-step session times, so values above one favor delta execution. Each pass aggregates 1,050 sessions as the median over seeds of the geometric mean over matrices. ``Native full'' is the maintained-support native CSC baseline and reports the median of three passes with the observed minimum and maximum in brackets. ``Best of 8'' uses the fastest of the eight full-recomputation implementations in each session, chosen after timing. The margin grows with $\rho$ because the cost of full recomputation grows with the input support.}
\label{tab:full}
\small\setlength{\tabcolsep}{4pt}
\begin{tabular}{@{}llrr@{}}
\toprule
Motion & $\rho$ & Native full / CSC & Best of 8 / CSC \\
\midrule
drift & 0     & 2.10 [2.10, 2.12] & 2.10 \\
drift & 0.002 & 2.29 [2.25, 2.34] & 2.29 \\
drift & 0.01  & 2.44 [2.39, 2.46] & 2.44 \\
drift & 0.05  & 3.05 [3.00, 3.07] & 3.04 \\
drift & 0.25  & 4.07 [4.07, 4.15] & 4.02 \\
jump  & 0     & 2.12 [2.11, 2.13] & 2.12 \\
jump  & 0.002 & 2.25 [2.21, 2.31] & 2.25 \\
jump  & 0.01  & 3.10 [3.00, 3.16] & 3.09 \\
jump  & 0.05  & 6.22 [5.88, 6.27] & 5.85 \\
jump  & 0.25  & 12.50 [11.89, 12.62] & 8.19 \\
\bottomrule
\end{tabular}
\end{table}

\UlPara{Tiled versus direct} Under the delta-product contract, we compare P3 with direct CSC, exact CSR rebuild and the two intermediate tiled variants on all 1,050 sessions.

\observation{obs:tiled}{The apparent benefit of schedule reuse depends on the baseline. The tiled executor outperforms exact CSR rebuild by up to 5.75$\times$, yet direct CSC outperforms it in 93\% of moving workloads.}

\begin{figure*}[tbp]
  \centering
  \includegraphics[width=\linewidth]{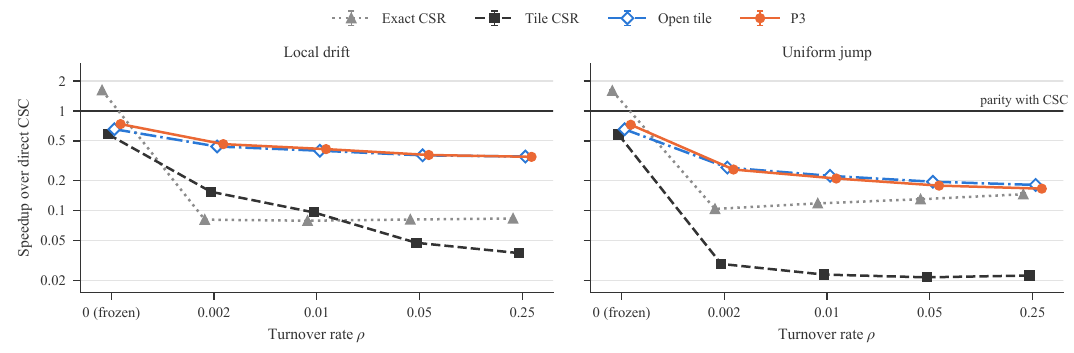}
  \caption{Steady-state speedup of each delta strategy over direct CSC in 400-step sessions under local drift (left) and uniform jump (right). The horizontal axis is the turnover rate $\rho$, and the logarithmic vertical axis shows speedup, so values above the line at one favor the strategy over CSC. Each marker is the median of three pass aggregates, and whiskers show the lowest and highest pass. Exact CSR rebuilds a row view whenever the dirty set changes, Tile CSR and Open tile retain partial reuse, and P3 retains operator preparation and caches up to eight sessions. P3 beats Exact CSR at every nonzero turnover rate, yet every tiled strategy stays below direct CSC at every turnover rate.}
  \label{fig:fivearm}
\end{figure*}

Fig.~\ref{fig:fivearm} reports steady-state speedups relative to direct CSC. Against exact CSR rebuild, P3 is 4.19--5.75$\times$ faster under drift and 1.14--2.51$\times$ faster under jump. Against direct CSC, the same P3 runs are 2.2--2.9$\times$ slower under drift and 3.9--6.0$\times$ slower under jump.

\begin{figure*}[tbp]
  \centering
  \includegraphics[width=\linewidth]{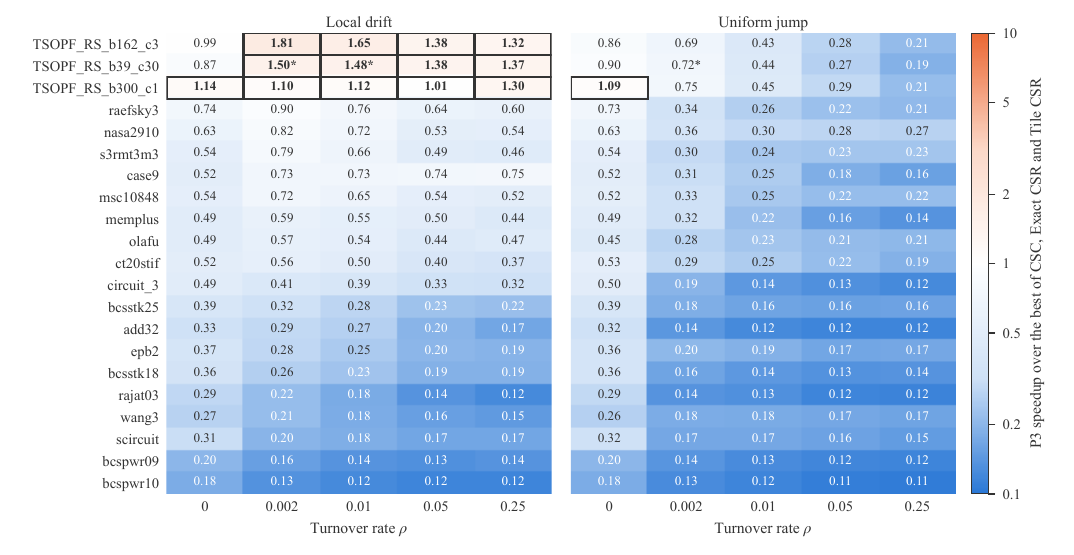}
  \caption{P3 speedup over the fastest simple strategy in each session (direct CSC, Exact CSR or Tile CSR, selected after timing) for every matrix, motion model and turnover rate. Each cell is the median of three pass values, and the color scale is logarithmic and centered at one. Bold, outlined values mark P3 advantages and blue cells mark P3 slowdowns. An asterisk marks a cell with one of five seeds omitted because its dirty set never changed. Rows are sorted by the drift $\rho=0.002$ column. P3 is faster only on the three TSOPF matrices, and under jump only for frozen \textnormal{\texttt{b300\_c1}}.}
  \label{fig:p3best}
\end{figure*}

Of 837 moving sessions, 834 produce the same faster strategy in all three passes. CSC is faster in 779 and P3 in 55. All 55 stable P3 advantages occur on the three TSOPF matrices, which come from transient-stability-constrained optimal power flow problems (Fig.~\ref{fig:p3best}). The ranking transfers across platforms. CSC is faster in 777 of 837 sessions on c7i.4xlarge and 723 of 837 on c7a.4xlarge. An independent replay of all 1,050 sessions agrees on the faster strategy in 99.88\% of the 821 sessions whose margin exceeds 10\%.

\UlPara{Source of the tiled advantage} We next examine why P3 is faster on the TSOPF matrices and on no other matrix.

\observation{obs:structure}{Tiled execution is faster only when tiles compress output row writes substantially. This is a structural property of the matrix. Within the TSOPF family it can be predicted before execution, but a cost model fitted to other families does not transfer reliably.}

\BfPara{Work decomposition} Table~\ref{tab:phases} instantiates the work model of Section~\ref{sec:background}. On \textnormal{\texttt{TSOPF\_RS\_\allowbreak b300\_c1}}, eight-column tiles cover 1.57$\times$ more entries than the active columns but reduce row writes by 4.78$\times$, which lowers application time from 687.3 to 549.5\,$\mu$s per step. On \textnormal{\texttt{scircuit}}, wider tiles increase both covered entries and row writes, and direct CSC remains 5.7$\times$ faster.

\begin{figure*}[t]
  \centering
  \includegraphics[width=.92\linewidth]{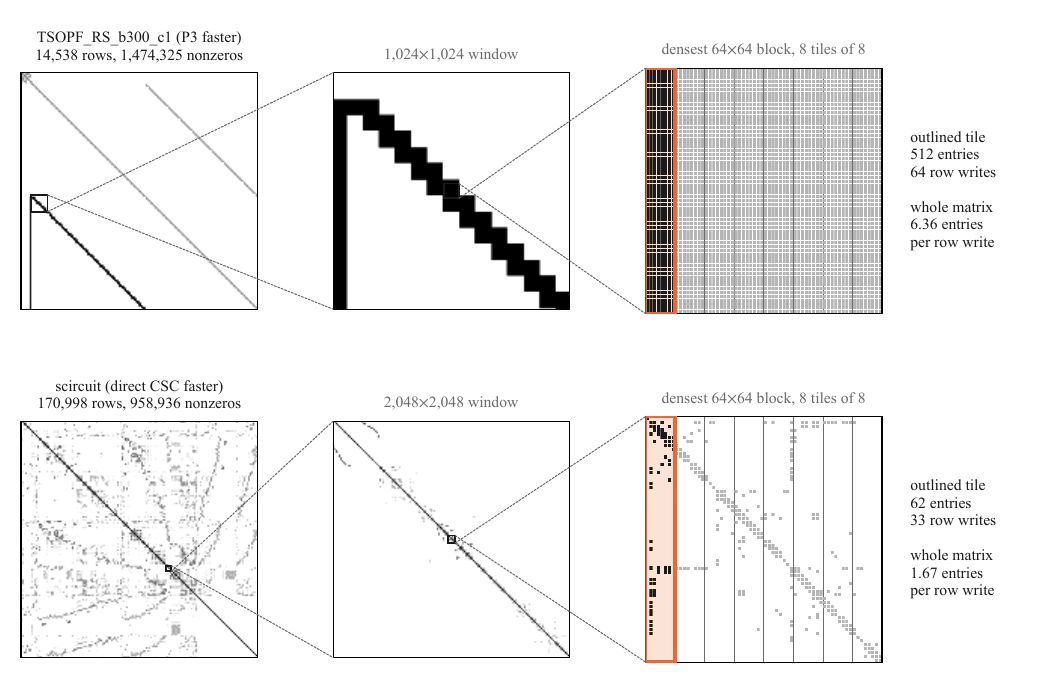}
  \caption{Structural source of the tiled advantage for \textnormal{\texttt{TSOPF\_RS\_b300\_c1}} (top), where P3 is faster, and \textnormal{\texttt{scircuit}} (bottom), where direct CSC is faster. Each row magnifies the matrix in three steps, from the full sparsity pattern (left, darker means more nonzeros) to a window around the densest region (middle) to the densest $64\times64$ block (right). The same selection rule applies to both matrices, and dotted lines connect each box to its magnification. Vertical lines in the right panels mark eight-column tiles. The orange tile holds the most entries, which are drawn in black. TSOPF columns that are neighbors share rows, so this tile holds 512 entries on only 64 rows and each row write combines eight entries. In \textnormal{\texttt{scircuit}} the entries are scattered, so the fullest tile holds 62 entries on 33 rows. Across each whole matrix, an eight-column tile combines 6.36 and 1.67 entries per row write.}
  \label{fig:zoom}
\end{figure*}

Fig.~\ref{fig:zoom} relates these counts to matrix structure. TSOPF matrices consist of dense blocks in which neighboring columns share rows. In the densest $64\times64$ block of \textnormal{\texttt{TSOPF\_RS\_\allowbreak b300\_c1}}, one eight-column tile holds 512 entries on 64 rows. The densest block of \textnormal{\texttt{scircuit}} yields 62 entries on 33 rows. Across the whole matrix, an eight-column tile combines 6.36 entries per row write for \textnormal{\texttt{TSOPF\_RS\_\allowbreak b300\_c1}} and 1.67 for \textnormal{\texttt{scircuit}}.

\begin{table}[t]
\centering
\caption{Per-step work of P3 by tile width $w$ under drift at $\rho=0.002$, from the eight-matrix panel (two seeds, three repeats). Active entries are the stored entries of the dirty columns, covered entries are all stored entries in the selected tiles, and writes are distinct (row, tile) pairs, all in thousands per step. Apply is the per-step application time in $\mu$s, and speedup is CSC time over P3 time (above one favors P3). On \textnormal{\texttt{b300\_c1}} wider tiles add covered entries but cut writes sharply, whereas on \textnormal{\texttt{scircuit}} they add both. This panel and the preregistered cohort of Table~\ref{tab:predict} (five seeds) are separate runs. Because \textnormal{\texttt{b300\_c1}} lies near parity, its speedup is 1.10 and 1.55 here against 0.86 and 1.12 there at widths 1 and 8.}
\label{tab:phases}
\small\setlength{\tabcolsep}{3pt}
\begin{tabular}{@{}lrrrrrr@{}}
\toprule
Matrix & $w$ & Active & Covered & Writes & Apply & Speedup \\
\midrule
b300\_c1 & 1  & 325.9 & 325.9 & 325.9 & 687.3 & 1.10 \\
b300\_c1 & 8  & 325.9 & 510.7 & 68.3  & 549.5 & 1.55 \\
b300\_c1 & 32 & 325.9 & 534.3 & 26.4  & 572.0 & 1.48 \\
scircuit & 1  & 1.9 & 1.9 & 1.9 & 42.1 & 0.14 \\
scircuit & 8  & 1.9 & 4.1 & 2.7 & 41.3 & 0.18 \\
scircuit & 32 & 1.9 & 8.7 & 5.4 & 59.6 & 0.21 \\
\bottomrule
\end{tabular}
\end{table}

\BfPara{Preregistered intervention} If write compression causes the tiled advantage, then a tile width of one column should remove it, and structurally similar matrices should show it. Before any intervention timing, we registered two predictions in a dated record committed to version control, and the \href{https://github.com/HorneSci/how-the-dirty-set-moves}{public verification artifact} includes this record with its SHA-256 hash. First, width one reverses the P3 advantage on at least one of the three original TSOPF matrices. Second, among all remaining TSOPF matrices of the same reduced-space (RS) formulation in SuiteSparse with 0.1--2 million nonzeros (\textnormal{\texttt{b162\_c1}}, \textnormal{\texttt{b162\_c4}}, \textnormal{\texttt{b39\_c7}}, \textnormal{\texttt{b39\_c19}}), width eight favors P3 on at least two and width one reverses at least one, in every repeat. The intervention fixes matrix order, dirty sets, values and output rows, and varies only the width over 1,260 sessions (seven matrices, two motions, two turnover rates, five seeds, three widths and three repeats) in three counterbalanced width orders.

Both predictions hold (Table~\ref{tab:predict}). Width eight favors P3 on all four held-out matrices by 1.75--1.94$\times$, and width one reverses all four, together with \textnormal{\texttt{b39\_c30}}, in every repeat. The registered predictions concern drift at $\rho=0.002$, the slice that Table~\ref{tab:predict} reports; the 1,260 sessions also cover drift at $\rho=0.05$ and jump at both rates. At drift $\rho=0.05$ both predictions also hold, as width eight favors P3 on all four held-out matrices by 1.30--1.61$\times$ and width one reverses all four. Under jump, P3 is slower at the median at every width on all seven matrices, so there is no P3 advantage to reverse. Because width also changes the inner loop, the intervention measures the joint effect of grouping, covered entries, row writes and code path; the phase counts attribute the dominant share to write compression.

\begin{table}[t]
\centering
\caption{Preregistered tile-width intervention on the three original TSOPF matrices and four held-out TSOPF matrices. Each entry is CSC time over P3 time (above one favors P3) at tile width $w$ under drift at $\rho=0.002$, reported as the median of three repeats of the five-seed median with the observed range in brackets. Matrix order, dirty sets, values and output rows are identical across widths. Both registered predictions hold. Width eight favors P3 on all four held-out matrices, and width one reverses all four and \textnormal{\texttt{b39\_c30}} in every repeat.}
\label{tab:predict}
\footnotesize\setlength{\tabcolsep}{2.2pt}
\begin{tabular}{@{}llrrr@{}}
\toprule
Matrix & Set & $w=1$ & $w=8$ & $w=32$ \\
\midrule
b300\_c1 & orig. & 0.86 [0.81, 1.02] & 1.12 [1.07, 1.45] & 1.09 [1.04, 1.59] \\
b162\_c3 & orig. & 0.88 [0.86, 1.09] & 1.97 [1.71, 2.18] & 2.31 [1.91, 2.59] \\
b39\_c30 & orig. & 0.72 [0.66, 0.81] & 1.65 [1.52, 2.02] & 1.93 [1.74, 2.85] \\
b162\_c1 & held-out & 0.91 [0.86, 0.91] & 1.75 [1.67, 1.77] & 2.02 [1.98, 2.03] \\
b162\_c4 & held-out & 0.87 [0.85, 0.93] & 1.78 [1.59, 1.95] & 2.04 [1.63, 2.52] \\
b39\_c7 & held-out & 0.91 [0.86, 0.92] & 1.94 [1.87, 1.94] & 2.18 [2.16, 2.32] \\
b39\_c19 & held-out & 0.90 [0.86, 0.95] & 1.88 [1.69, 2.13] & 2.29 [1.83, 2.50] \\
\bottomrule
\end{tabular}
\end{table}

\BfPara{Predictability} We fit a linear cost model with nonnegative coefficients to the work-model counts. To test it on unseen structure, each matrix family (the TSOPF group, or a single other matrix) is excluded from training when its own observations are predicted. We report \emph{regret}, the cost of the selected strategy divided by the cost of the faster one. The model selects the faster strategy in 203 of 240 observations, against 163 for always selecting CSC. Its mean regret on TSOPF is 1.001, against 1.670 for CSC. With families weighted equally, however, its mean regret is 1.297 against 1.143 for CSC. The model is accurate where tiling is faster and costly elsewhere.

\UlPara{Reuse lifetimes} We now vary each lifetime of reuse in isolation while the kernel and input streams stay fixed.

\observation{obs:lifetimes}{Operator preparation carries the benefit of reuse. Session capacity beyond the current session yields almost no additional hits under realistic motion, and setup cost favors direct CSC for short epochs.}

\BfPara{Operator preparation}
Nine fresh processes repeat all 1,050 primary sessions with persistent tile preparation enabled and disabled; cache capacity, tile groups, values and kernel remain fixed. Retention is faster in all three repeats on 823 workloads and slower on 163, with 64 mixed. Every matrix benefits at the median (1.79--4.05$\times$). The geometric mean over matrices is 2.66$\times$, or 1.72$\times$ with setup charged. A transfer sweep over batch widths $B\in\{1,8,32\}$ and dirty-set targets of 64, 256 and 1,024 rows shows a benefit in all nine settings (1.10--2.58$\times$ with setup charged). 

\BfPara{Session capacity} Holding the kernel and tile width fixed, we vary the session cache from one entry to eight (Table~\ref{tab:cache}). A capacity of one already reuses the current session whenever consecutive steps share a tile set, which gives a 50.9\% hit rate under drift. Capacity eight adds reuse of earlier sessions, but it raises the hit rate only to 52.4\% under drift (a median of two additional hits per 400 steps) and leaves it at zero under jump, and it is slightly slower in both cases. Only a synthetic four-set cycle, the access pattern that caching targets, produces a benefit (396 additional hits, 1.084$\times$). The larger cache also raises peak buffer memory from 0.36--20.87\,MiB to 1.60--93.93\,MiB.

\begin{table}[t]
\centering
\caption{Session capacity with the same P3 kernel and tile width eight, over eight matrices and 400-step sessions. Pairs counts paired observations, and ratio is the median of P3 time with one cached session over P3 time with eight (above one favors eight). Drift and jump pool turnover rates $\rho=0.002$ and $0.05$ over two seeds. Hits give the share of steps that reuse a cached session at capacity one and at capacity eight; the drift value of about one half reflects tile sets that stay unchanged in about 81\% of transitions at $\rho=0.002$ but about 6\% at $\rho=0.05$ (Fig.~\ref{fig:workload}), extra hits are the median number of additional reuses per session at capacity eight, and the last column counts pairs in which capacity eight is faster. Capacity one already reuses the current session in half of the drift steps, and the extra capacity improves performance only in the synthetic four-set cycle.}
\label{tab:cache}
\small\setlength{\tabcolsep}{4pt}
\begin{tabular}{@{}lrrrrr@{}}
\toprule
Motion & Pairs & Ratio & Hits 1/8 (\%) & Extra hits & Cap.-8 faster \\
\midrule
drift  & 96 & 0.985 & 50.9/52.4 & 2   & 41 \\
jump   & 96 & 0.931 & 0.0/0.0 & 0   & 10 \\
cycle  & 48 & 1.084 & 0.0/99.0 & 396 & 47 \\
\bottomrule
\end{tabular}
\end{table}

\BfPara{Setup amortization} P3 requires more setup than CSC. On \textnormal{\texttt{TSOPF\_RS\_\allowbreak b300\_c1}}, setup costs 256.3\,ms for P3 against 153.9\,ms for CSC (Table~\ref{tab:setup}). A steady-state P3 advantage therefore becomes a disadvantage for short epochs (0.645 after 16 steps) and becomes a gain only between 64 and 400 steps. On the remaining matrices, P3 remains slower after 400 steps.
Across the transfer sweep, CSC is faster in 26,715 of 28,350 setup-inclusive observations.

\begin{table}[t]
\centering
\caption{Setup cost and setup-inclusive comparison of CSC and P3 under drift at $\rho=0.002$ (two seeds, medians of three AWS repeats). Setup is the one-time construction of each operator from identical CSR inputs, in ms. The last three columns give total CSC time over total P3 time, setup and output resets included, after 16, 64 and 400 steps (above one favors P3). P3 needs roughly twice the setup time of CSC, so even on \textnormal{\texttt{b300\_c1}}, where it is faster in steady state, it becomes profitable only between 64 and 400 steps.}
\label{tab:setup}
\small\setlength{\tabcolsep}{4pt}
\begin{tabular}{@{}lrrrrr@{}}
\toprule
 & \multicolumn{2}{c}{Setup (ms)} & \multicolumn{3}{c}{CSC/P3 after $L$ steps} \\
Matrix & CSC & P3 & 16 & 64 & 400 \\
\midrule
b300\_c1 & 153.9 & 256.3 & 0.645 & 0.778 & 1.093 \\
scircuit & 88.9 & 170.6 & 0.517 & 0.529 & 0.694 \\
ct20stif & 231.4 & 452.5 & 0.520 & 0.526 & 0.571 \\
add32 & 1.3 & 2.7 & 0.471 & 0.492 & 0.531 \\
\bottomrule
\end{tabular}
\end{table}

\UlPara{End to end} Finally, we charge the costs outside the kernel in an application pipeline, in online strategy selection and in parallel execution.

\observation{obs:e2e}{Kernel-level gains shrink once application pipelines, online selection or communication are charged. Delta execution retains its advantage on multicore hosts; across hosts, communication dominates at small update sizes.}

\BfPara{Recommendation pipeline} We replay MovieLens 100K~\cite{harper2015,movielens100k}. The first 80,000 ratings train a fixed 1,682-item cosine-similarity operator with 51,693 nonzeros. The remaining 2,860 ratings arrive in 179 batches for eight users, and each batch must produce complete scores and top-ten recommendations that match a SciPy reference. With 32-rating profiles, direct CSC is 2.50$\times$ faster than maintained-support full recomputation at the kernel level, but the pipeline ratio is 0.867$\times$ once profile maintenance, construction of the compact update, reads and ranking are charged (Table~\ref{tab:app}). With complete histories, an 8.16$\times$ kernel advantage becomes 1.141$\times$.
In a second replay of air-quality sensor windows, specialized running sums outperform every sparse strategy in all 12 replays.

\begin{table}[t]
\centering
\caption{MovieLens recommendation replay. Eight users receive 2,860 held-out ratings in 179 batches, and after each batch both strategies return complete scores and top-ten recommendations that match a SciPy reference. Entries are maintained-support full-recomputation cost over direct CSC cost for the multiplication kernel alone, the complete pipeline and the pipeline with setup (above one favors delta execution), as medians of three repeats. Profiles keep either the complete rating history or the 32 most recent ratings. With recent-32 profiles, a 2.50$\times$ kernel advantage disappears over the full pipeline because profile upkeep, construction of the compact update, reads and ranking dominate.}
\label{tab:app}
\small\setlength{\tabcolsep}{4pt}
\begin{tabular}{@{}lrrr@{}}
\toprule
Profile & Kernel & Pipeline & Epoch \\
\midrule
Complete history & 8.16 & 1.141 & 1.126 \\
Recent 32 & 2.50 & 0.867 & 0.878 \\
\bottomrule
\end{tabular}
\end{table}

\BfPara{Online selection}
Deployed as a causal selector with family-held-out coefficients, the cost model requires 39.51\,ms of metadata construction and 53.95\,$\mu$s per decision. With these costs charged, always selecting CSC is faster (CSC/selector 0.780), and CSC is faster in 1,135 of 1,200 observations.

\BfPara{Multicore execution} On c7i.4xlarge with one to eight physical cores and OpenMP~\cite{openmp} (six matrices, 216 paired measurements per strategy), full CSR with output rows partitioned across workers (row-partitioned) scales best and reaches 5.45$\times$ at eight workers, against 3.19$\times$ for row-partitioned CSC and 2.17$\times$ for row-partitioned P3. Delta work per step is small and unevenly distributed across row partitions. Nevertheless, at eight workers the fastest delta strategy outperforms the fastest full strategy in every observation of all 36 conditions, by 1.68--17.3$\times$ in steady state and 1.25--2.15$\times$ setup-inclusive (Fig.~\ref{fig:parallel}a).

\BfPara{Distributed execution} We partition rows across processes (ranks) on up to two hosts and communicate with Gloo~\cite{gloo}. One rank on one host gives the lowest cost for CSC, P3 and oneMKL (Fig.~\ref{fig:parallel}b). At four ranks on two hosts, the median speedups relative to one rank are 0.081 for CSC, 0.165 for P3 and 0.544 for oneMKL. Computation accounts for 8.8\% of P3's shard-ready time, that is, the time until every rank holds its own output rows. A control that performs the same broadcasts, readiness checks and gathers without arithmetic costs 200.75\,ms per 64-step session, against 222.62\,ms for P3. At these update sizes, communication rather than computation determines distributed cost.

\IfFileExists{figs/fig_parallel.pdf}{%
\begin{figure}[t]
  \centering
  \includegraphics[width=\linewidth]{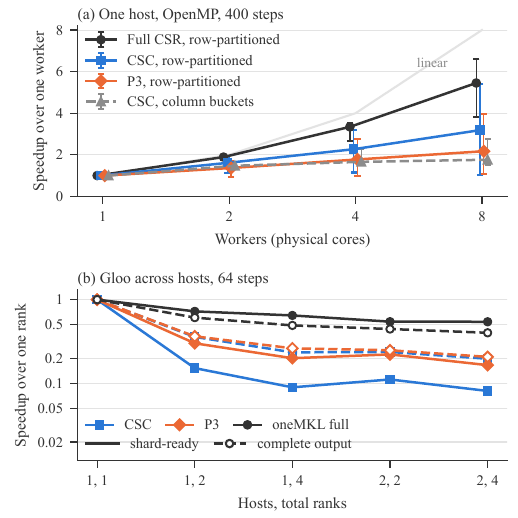}
  \caption{Parallel and distributed scaling for six matrices, relative to one worker or one rank. (a) Shared-memory runs on c7i.4xlarge with OpenMP and 400-step sessions. Markers show medians and bars show interquartile ranges of 216 paired ratios per strategy and worker count, and the gray line marks linear speedup. Row-partitioned full CSR scales best, while the delta strategies scale less because their per-step work is small and uneven. (b) Gloo runs on up to two hosts with 64-step sessions, on a logarithmic axis whose horizontal labels give hosts and total ranks. Solid lines stop when every rank holds its own output rows (shard-ready), and dashed lines include gathering the complete output at the root. Every strategy slows down as ranks are added because communication dominates.}
  \label{fig:parallel}
\end{figure}%
}{}

\UlPara{Synthesis} Delta execution is the first-order optimization. Schedule reuse is a second-order optimization that is profitable only when tiles compress row writes and the epoch is long enough to amortize preparation.

\section{Design Guidelines\aicite}
\label{sec:guidelines}

\BfPara{G1. Adopt delta execution and compare it with full recomputation} Studies of incremental kernels should include full recomputation and direct CSC as baselines; as Observation~\ref{obs:tiled} shows, conclusions can otherwise reverse.

\BfPara{G2. Use direct CSC as the default} Direct CSC is the most reliable strategy across motion models, turnover rates and platforms.

\BfPara{G3. Retain operator preparation; enlarge session caches only for recurring workloads} Per-tile preparation should persist for the lifetime of the operator; extra session capacity is beneficial only in periodic workloads and otherwise costs up to 4.5$\times$ more buffer memory.

\BfPara{G4. Select tiling per operator at setup time} Tile operators whose tiles combine many entries per row write, such as the TSOPF family, and only for epochs of hundreds of steps.

\BfPara{G5. Charge complete pipeline and communication costs} Profile maintenance, top-$k$ selection and delivery can eliminate a 2.5$\times$ kernel advantage, and small distributed updates are dominated by communication latency.

\section{Related Work\aicite}
\label{sec:related}

\BfPara{Sparse kernels for sparse inputs} Work-efficient SpMSpV merges selected CSC columns into output buckets~\cite{azad2017}. GPU and HPC designs rely on sort and merge steps or on fine-grained parallelism~\cite{yang2015,chen2022}, and TileSpMSpV applies tiled push and pull kernels~\cite{ji2022tile}.  DB-SpMSpV, Adaptive SpMV/SpMSpV and HAM-SpMSpV appear in Table~\ref{tab:closest}~\cite{cong2026db,li2021adaptive,xu2024ham}. These studies time kernels, typically along breadth-first search frontiers, and measure conversion separately. We instead treat the lifetime of prepared state as the variable under study and charge setup, state and delivery.

\BfPara{Inspection and code generation} Inspector-executor frameworks such as OSKI amortize analysis across calls~\cite{vuduc2005oski,strout2018}. TACO generates sparse iteration code~\cite{kjolstad2017}, and reordering improves locality~\cite{akbudak2013}. GraphBLAS implementations express graph computations as sparse linear algebra~\cite{yang2022graphblast,davis2019}, and sequence SpMV pipelines repeated products with dense vectors~\cite{zhang2023sspmv}. These techniques assume a stable access pattern; in delta-input SpMM the pattern changes at every step.

\BfPara{Incremental computation} Incremental view maintenance~\cite{koch2014}, self-adjusting computation~\cite{acar2002}, differential dataflow~\cite{mcsherry2013}, systems for evolving graphs~\cite{mariappan2021} and DeltaCNN~\cite{parger2022} retain state to avoid recomputation. They motivate delta execution but do not examine schedule reuse within the sparse kernel. Evolving-graph systems such as KickStarter, GraphBolt and DZiG update query results incrementally~\cite{vora2017kickstarter,mariappan2019graphbolt,mariappan2021}, and UVVs identifies vertex values that remain unchanged across snapshots to shrink per-snapshot work~\cite{afarin2026uvvs}. These systems decide which values to recompute; our study concerns how the sparse kernel that performs the update should reuse its schedule. Table~\ref{tab:closest} contrasts these kernels with this study. Distributed SpMM work addresses communication volume and overlap~\cite{koanantakool2016,block2024twoface}; our distributed results show that communication also dominates small incremental updates.

\section{Threats to Validity\aicite}
\label{sec:validity}
Several conclusions depend on assumptions that we state explicitly. \ding{182}~The tiled results come from one executor design, P3. Designs with different preparation or GPU execution, such as DB-SpMSpV, may shift the break-even point for tiling. \ding{183}~The defaults $B=8$, $w=8$ and a capacity of eight shape the magnitudes; the preparation benefit shrinks as $B$ grows, and $w$ changes which matrices favor tiling. \ding{184}~GPUs and other platforms with costlier output writes may favor tiling more often than our CPU results suggest. \ding{185}~Support motion is synthetic except for the two application replays, and periodic workloads may benefit from caching. No run disables session reuse entirely, so the capacity study measures only reuse beyond the current session. \ding{186}~The corpus is a preselected sample of 21 matrices, and all stable P3 advantages come from one family. The held-out intervention establishes transfer within that family, not across structural classes. \ding{187}~Tile width changes grouping, covered entries, row writes and code path jointly, so the intervention identifies their combined effect. Primary timings use a single core, and the multicore and distributed studies use six matrices and a small network configuration.

\section{Conclusion\aicite} Delta execution consistently outperforms full recomputation, but schedule reuse seldom outperforms direct CSC. Its benefit comes from operator preparation, and tiled execution is faster only on operators whose structure compresses row writes. How the dirty set moves, and how the operator groups it, determines whether schedule reuse is profitable.

\section*{Acknowledgements} OpenAI Codex~\cite{openaicodex} and Anthropic Claude~\cite{anthropicclaude} assisted with drafting and revising the text. 

\bibliographystyle{IEEEtran}
\bibliography{refs}
\end{document}